\documentclass[prb,twocolumn,showpacs,floatfix,amsmath,amssymb]{revtex4}

\usepackage{graphicx}
\usepackage{dcolumn}
\usepackage{bm}

\begin{document}


\title{Supercurrent-phase relationship of a Nb/InAs(2DES)/Nb
Josephson junction in overlapping geometry}

\author{Mark Ebel}
 \email{ebel@physik.uni-hamburg.de}
\author{Christian Busch}%
\altaffiliation[Now at ]{Forschungszentrum J\"{u}lich, Institut f\"{u}r Plasmaphysik, D-52425 J\"{u}lich, Germany}
\author{Ulrich Merkt}%

\affiliation{Institut f\"{u}r Angewandte Physik, Universit\"{a}t
Hamburg, Jungiusstr.~11, D-20355 Hamburg, Germany}

\author{Miroslav Grajcar}%
\altaffiliation[Also at ]{Department of Solid State Physics,
Comenius University, Mlynsk\'{a} Dolina F2, 842 48 Bratislava,
Slovakia}

\author{Tom\'{a}\v{s} Plecenik}%
 \altaffiliation[Also at ]{Department of Solid State Physics,
Comenius University, Mlynsk\'{a} Dolina F2, 842 48 Bratislava,
Slovakia}

\author{Evgeni Il'ichev}

\affiliation{Department of Cryoelectronics, Institute for Physical
High Technology, P.O. Box 100239, D-07702
Jena, Germany}%

\date{\today}

\begin{abstract}
Superconductor/normal conductor/superconductor (SNS) Josephson
junctions with highly transparent interfaces are predicted to show
significant deviations from sinusoidal supercurrent-phase
relationships (CPR) at low temperatures. We investigate
experimentally the CPR of a ballistic Nb/InAs(2DES)/Nb junction in
the temperature range from 1.3 K to 9 K using a modified
Rifkin-Deaver method. The CPR is obtained from the inductance of
the phase-biased junction. Transport measurements complement the
investigation. At low temperatures, substantial deviations of the
CPR from conventional tunnel-junction behavior have been observed.
A theoretical model yielding good agreement to the data is
presented.
\end{abstract}


\maketitle

\section{Introduction}

A fundamental item in the theoretical modeling of a Josephson
junction is the dependency of the supercurrent $I_S$ flowing
through it on the phase difference $\varphi$ across the junction.
Golubov, Kupriyanov and Il'ichev published a recent review on the
subject.\cite{golubov04} As this relation is necessarily
$2\pi$-periodic and odd,\cite{likharev79} it can be expressed as
the following Fourier series (with the critical current $I_C$):
\begin{equation}
I_S(\varphi)=I_Cf(\varphi)=\sum_n{I_n\sin(n\varphi)}.
\label{CPR_fourier}
\end{equation}
$I_S(\varphi)$ is known as (super-)current-phase relationship or
CPR, the dimensionless term $f(\varphi)$ as normalized CPR. For
vanishing transparency (e.g.~a tunnel junction), only $I_1$ is
relevant, reducing Eq.~\ref{CPR_fourier} to the well-known dc
Josephson\cite{josephson62} equation
\begin{equation}
 I_S(\varphi)=I_1\sin\varphi=I_C\sin\varphi.
 \label{dcJosephson}
\end{equation}
Junctions with direct (i.e.~non-tunnel) conductivity, such as the
Nb/InAs(2DES)/Nb weak links used in this work, are predicted to
show more complex behavior and a non-sinusoidal CPR due to higher
order processes of charge transport such as multiple Andreev
reflections.\cite{ko-2,likharev79,golubov04} Indications of a
significant non-harmonic term in our junctions have been found in
microwave measurements.\cite{baars03,baars03.2} We are interested
in experimental access to the CPR as a way to test and improve the
theoretical models describing our junctions, since the conduction
mechanisms in these devices are still not fully understood. The
measurements deliver not only the dependency $f(\varphi)$, but
also $I_C$, thus offering an independent comparison to transport
measurements.

\section{Theory}

The first extensive theoretical works concerning high transparency
superconducting weak links were published by Kulik and
Omel'yanchuk for a short quasiclassical point contact in the dirty
limit\cite{ko-1} $l\ll\xi_0$ (with mean free path $l$ and
coherence length $\xi_0$)  and for the clean limit\cite{ko-2}
$l\gg\xi_0$, predicting a CPR $f(\varphi)$ changing from a
sinusoidal curve to a saw tooth shape for high transparencies and
low temperatures. These idealized model systems serve to
understand the basic mechanisms leading to a deviation of the CPR
from the well-known Josephson relation (Eq.~\ref{dcJosephson}),
which we encounter also in complex real-world junctions, though
mixed with secondary processes. Mechanically controlled
break-junctions\cite{muller94} come closest to the assumptions of
the theory of Kulik and Omel'yanchuk. CPR measurements by a
flux-detecting method in this system\cite{koops96} agreed with the
predicted changes in the position of the maximum and curve shape.
In contrast to these systems, our SNS junctions exhibit a finite
length, some scattering of electrons in the 2DES, and an area of
induced superconductivity around the electrodes caused by the
proximity effect. To take these properties into account and to
describe the non-sinusoidal CPR in our experiments, we used the
scattering matrix formalism\cite{brouwer97} to derive a formula
for the Josephson current through a double-barrier
structure:\cite{grajcar02}
\begin{widetext}
\begin{equation}\frac{eIR_0}{\pi
k_BT_C}=\frac{16T}{T_C} \sum_{n=0}^{N-1}\sum_{m=0}^{\infty}
\frac{-A^2|S_{12}|^2\sin(\varphi)}
{1-2A^2(|S_{12}|^2\cos(\varphi)+|S_{11}|^2)+A^4|S_{11}^2+S_{12}^2|^2},
\label{eq:In}
\end{equation}
\end{widetext}
where $R_0=h/e^2$ is the resistance quantum, $A$ is the
coefficient of the Andreev reflection at the InAs(2DEG)/InAs
interface and $S_{ij}$ are the elements of the scattering matrix
for normal reflection in a symmetric double barrier
junction\cite{belogolovski99}
$S_{11}=|r|+|r||t|^2p_n^2/(1+|r|^2p_n^2)$,
$S_{12}=|t|^2p_n/(1+|r|^2p_n^2)$. Here $|r|^2, |t|^2$ are the
reflection and transmission probabilities of the left and
right-hand barrier, $p_n=\exp(ik_{Fn}a-\omega_m a/\hbar v_{Fn})$,
where $\omega_m=(2m+1)\pi k_BT$ are Matsubara frequencies,
$k_{Fn}$ and $v_{Fn}$ are components normal to the barriers of the
wave vector and Fermi velocity, respectively, of the $n$-th
transverse mode to the barriers and $a$ is the distance between
them. Following the BTK approach,\cite{blonder82} the reflection
and transmission coefficients can be written in the form
\begin{equation}
 |t|^2=1-|r|^2=\frac{1-(n/N)^2}{(1-(n/N)^2)(\eta+1)^2/4\eta+Z^2}
\label{eq:t}
\end{equation}
where $Z$ is the dimensionless potential barrier strength and
$\eta =v_{Fs}/v_{Fn}$ is the Fermi velocity mismatch.

In order to calculate the Josephson current from Eq.~\ref{eq:In},
one must determine the Andreev reflection coefficient $A$ for the
InAs(2DEG)/InAs interface. Since superconductivity is induced in
InAs by the proximity effect, the coefficient $A$ may be written
as \cite{aminov96} $A=iF/(1+G)$, where $F$ and $G$ are Green
functions in the inversion layer of InAs. Due to the low electron
density of the inversion layer the suppression of the pair
potential in Nb can be neglected and $F, G$ can be expressed by
the McMillan equations\cite{golubov95}
$G=\omega/\sqrt{\omega^2+\Phi^2}$,
$F=\Phi/\sqrt{\omega^2+\Phi^2}$, and
$\Phi=\tilde{\Delta}/(1+\gamma_B\sqrt{\tilde{\omega}^2+\tilde{\Delta}^2})$,
where $\gamma_B$ is a dimensionless parameter characterizing the
transparency between Nb and InAs, $\tilde{\Delta}=\Delta/\pi
k_BT_C$, $\tilde{\omega}=\omega/\pi k_BT_C$, $\Delta$ is the
superconducting energy gap of bulk Nb, and $T_C$ its critical
temperature. Both the critical temperature and the energy gap of
the Nb can be suppressed near the interface because of disorder
\cite{Belitz94} but $\tilde{\Delta}\approx 0.6$ remains constant.
\cite{Smith95} Since $T_C$ can be determined from temperature
measurement, there are only two free fitting parameters in the
model, the carrier density $n_s$ and  $\gamma_B$ (we assume $Z=0$,
i.e.\ no real barrier). The Fermi velocity $v_F$ and Fermi wave
vector $k_F$ are calculated for a given value of $n_s$. Since the
normal resistance depends on $n_s$ as well, the $n_s$ obtained
from the fit can be verified comparing  the theoretical and
experimental value of the resistance.

\begin{figure}[t]
\includegraphics[width=0.45\textwidth,keepaspectratio]{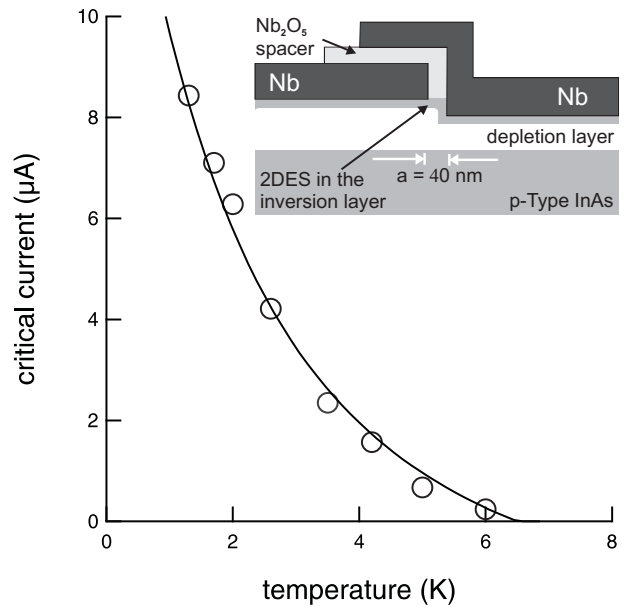}
\caption{\label{qoverlap}Temperature dependence of the critical
current. Circles are experimental data, the solid line presents a
fit using the model described in the text. The inset shows a
schematic cross section of the Josephson
junction.\label{1095-21a_IcT}}
\end{figure}

\section{Sample preparation}

The superconductor used in our SNS junctions is a Nb thin film,
the normal conductor is the two-dimensional electron system (2DES)
that forms as naturally occurring inversion layer at the surface
of bulk p-type InAs. The InAs surface is cleaned in situ using
low-energy Ar etching prior to the deposition of a 100 nm thick Nb
film by magnetron sputtering, yielding the highly transparent
interfaces required to observe deviations from sinusoidal CPR. The
structures are patterned using standard optical and electron-beam
lithography. We are using an overlapping sample geometry (see
inset of Fig.~\ref{qoverlap}), where we can set the electrode
separation $a$ with nm accuracy using anodic oxidation to grow the
insulating Nb$_2$O$_5$ interlayer. At a typical electron density
of $n_s=1.2\times 10^{12}$~cm$^{-2}$ and mobility of $\mu\approx
10^4 $ cm$^2$/Vs, the mean free path of $l=240$ nm is much longer
than the electrode separation $a=40$ nm. From the dependence of
the critical current on the channel length, we estimate the
coherence length $\xi_N \approx 145$ nm at 1.8 K. Thus we have a
ballistic ($a\ll l$) superconducting weak link in the short limit
($a\ll\xi_N$). More details have been published
by~\citet{chrestin97.1}

The use of the native 2DES on p-type InAs is not without
difficulties, as it forms on every surface of the crystal. Thus it
offers parallel conduction paths, resulting in significant bypass
currents around the junction, which reduce the normal resistance
$R_N$.

\section{CPR Measurement Technique}
\begin{figure}[t]%
   \includegraphics[width=0.45\textwidth,keepaspectratio]{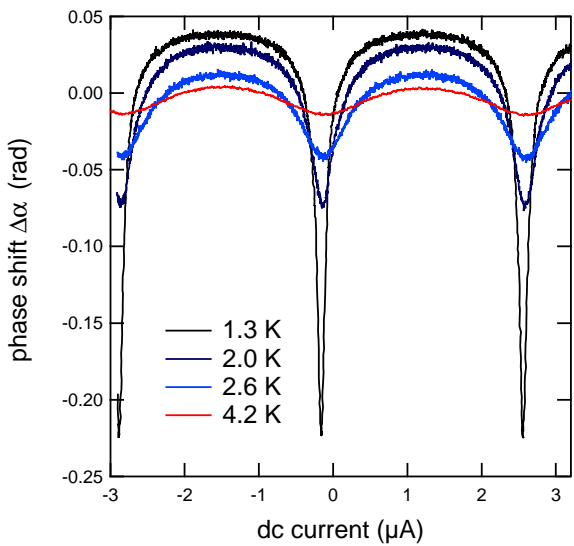}%
    \caption{ \label{cprrawdata}Measured phase shift at various temperatures.}
\end{figure}

The CPR of the Josephson junction is determined by an inductive rf
frequency method requiring no galvanic contacts to the sample,
thus reducing noise. This technique is based on the work of Rifkin
and Deaver.\cite{rifkin76} The junction to be investigated is
incorporated into a superconducting loop. This loop forms an rf
SQUID of inductance $L$, which is coupled inductively to a high
quality tank circuit in resonance. The phase difference $\varphi$
across the junction can be biased by an external magnetic flux
$\Phi_e$ via a dc current $I_\mathrm{dc}$. Changes in the
impedance of the coupled system are measured and can be used to
reconstruct the CPR. Further details are given in
Ref.~\onlinecite{golubov04}. This method requires low critical
currents $I_C$ of the junction and low inductances $L$ of the
SQUID washer, as the SQUID enters the hysteretic regime when
$\beta f'(\varphi)>1$, where $\beta=2\pi LI_C/\Phi_0$ is the
normalized critical current and $f'(\varphi)=
df(\varphi)/d\varphi$. In the hysteretic mode the internal
magnetic flux as a function of the external one becomes
multivalued, preventing us to reconstruct the CPR in the complete
phase range $(0,2\pi)$. The challenge is to reduce $I_C$ and $L$
while maintaining excellent interface transparency, large $I_CR_N$
products, and sufficient coupling to the tank circuit. We achieve
this by reducing the geometric dimensions of the junction and the
loop and by complex washer designs. The presented junction is
connected to six SQUID washers in parallel with an integrated flux
transformer, for an inductance of $L=17$~pH and optimized coupling
to the tank circuit. A similar transformer is described and
depicted in Ref.~\onlinecite{golubov04}.

\section{Results}

The experiments are performed on an overlapping Josephson junction
of width $w = 1~\mu$m and electrode separation $a = 40$ nm.
Transport measurements at 1.8~K result in $I_C=3.8~\mu$A,
$R_N=58~\Omega$, $I_CR_N=220~\mu$V, and show clearly developed
subharmonic gap structures (SGS) in the differential resistance of
the junction. This indicates high interface quality and
transparency of the junctions.

\begin{figure}[t]%
\includegraphics[width=0.45\textwidth,keepaspectratio]{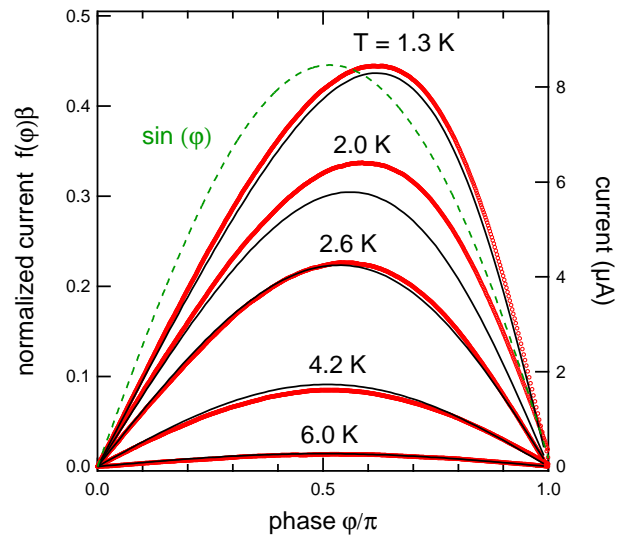}%
\caption{\label{overlapcpr}CPR (red) at
    temperatures between 1.3 K and 6.0 K,
     reconstructed from the data in Fig.~\ref{cprrawdata}.
     The solid lines are a fit according to the model described in the text. For comparison,
     a pure sine curve is included.}%
\end{figure}

Figure~\ref{cprrawdata} shows the recorded phase shift $\alpha$ in
the tank circuit as a function of the applied quasi-dc current
$I_\mathrm{dc}$ ramping the external flux $\Phi_e$ through the
SQUID loop. This signal was averaged for 20 periods to reduce
noise and then used to reconstruct the CPR depicted  in
Fig.~\ref{overlapcpr}. For decreasing temperatures, gradual
deviation of the CPR from conventional sinusoidal behavior towards
a sawtooth-shaped curve is observed. The position of maximum
current is shifted to $0.63\pi$ at 1.3 K. At the same time, the
critical current increases substantially, so that at some
temperature the SQUID enters the hysteretic regime as $\beta
f'(\varphi)$ approaches unity. In that case we are no longer able
to reconstruct the CPR at lower temperatures.

The $I_C-T$ dependence was fitted by the model described above
using the least squares method with two fitting parameters $n_s$
and $\gamma_B$. The other parameters used in the calculation were
determined independently as critical temperature $T_C=6.5$  K,
width $w=1.0~\mu$m, electrode separation $a=40$ nm, interface
parameter $Z=0$, and Fermi velocity mismatch $\eta=0.93$. The
fitting procedure yields the parameters $n_s=3.75\times10^{12}$
cm$^{-2}$ and $\gamma_B=2.4$ and good agreement between
theoretical and experimental data (see Figs.~\ref{1095-21a_IcT}
and \ref{overlapcpr}). Note that the model fits the current-phase
characteristic quite well for the parameters determined from the
temperature dependence of the critical current. Only the curve
taken at $T\approx 2.0$ K exhibits some deviation from the
theoretical one. However, the temperature dependence of $I_C(T)$
is rather steep at low temperatures and a small difference between
the temperature of the thermometer and the sample could be
responsible. From the above parameters we calculate the junction
resistance of $R_N=83~\Omega$. Transport measurements yield a
slightly lower value of $58~\Omega$, but as we measure a parallel
connection between substrate and junction, a higher junction
resistance is to be expected. A realistic substrate resistance of
$180~\Omega$ produces the measured value of $58~\Omega$. The
interface parameters indicate a highly transparent junction. The
assumed carrier density is about three times the result
$n_s=1.2\times10^{12}$ cm$^{-2}$ from Shubnikov-de Haas
measurements  on comparable samples. Nevertheless, $n_s$ as
determined from our model is consistent with the smaller
experimental value of the junction resistance. A possible
explanation for both effects is an effective widening of the
contact due to the proximity effect in the surrounding 2DES.

\section{Conclusion}

We have successfully measured the current-phase relationship (CPR)
of a Nb/InAs(2DES)/Nb Josephson junction in dependence of
temperature. At low temperatures, substantial deviations of the
CPR from a sinusoidal behavior towards a saw tooth shape are
observed. This is in qualitative agreement with predictions of the
Kulik-Omel'yanchuk theory for highly transparent SNS junctions and
the measurements on Josephson field effect transistors presented
in Ref.~\onlinecite{grajcar02}. A model yielding good quantitative
agreement to the results is presented. Transport measurements
support the results gained by phase sensitive measurements.

\begin{acknowledgments}
We thank V. Zakosarenko for help with the washer design and
gratefully acknowledge financial support by the European Science
Foundation via the PiShift program and by the Deutsche
Forschungsgemeinschaft via the SFB 508 Quantenmaterialien. M.G.
wants to acknowledge partial support by Grant Nos. VEGA 1/9177/02
and APVT-20-021602.
\end{acknowledgments}

\appendix

\end{document}